\documentclass[twocolumn,preprintnumbers,amsmath,amssymb,floatfix,superscriptaddress,nofootinbib]{revtex4}

\usepackage{amsmath,hyperref,amssymb}
\usepackage{graphicx}
\usepackage{dcolumn}
\usepackage{bm}
\usepackage{verbatim}
\usepackage{tabularx}
\usepackage{slashed}
\usepackage{cancel}
\usepackage[boxsize=2em,aligntableaux=center]{ytableau}
\usepackage{float}

\def\be{\begin{equation}}
\def\ee{\end{equation}}
\def\bea{\begin{eqnarray}}
\def\eea{\end{eqnarray}}

\newcommand{\ZZ}{\mathbb{Z}}
\newcommand{\LL}{\mathcal{L}}
\newcommand{\OO}{\mathcal{O}}
\newcommand{\DD}{\mathbb{D}}

\begin{document}

\vspace*{-30mm}

\title{Solving the Hierarchy Problem Discretely}
\author{Anson Hook}

\affiliation{Maryland Center for Fundamental Physics, Department of Physics\\
University of Maryland, College Park, MD 20742.}

\vspace*{1cm}

\begin{abstract} 

We present a new solution to the Hierarchy Problem utilizing non-linearly realized discrete symmetries.
The cancelations occur due to
a discrete symmetry that is realized as a shift symmetry on the scalar and as an exchange symmetry on 
the particles with which the scalar interacts.
We show how this mechanism can be used to solve the Little Hierarchy Problem as well as give rise to light axions.

\end{abstract}

\maketitle

\section{Introduction}

In this article, we describe a new mechanism for obtaining a scalar that is parametrically lighter than the UV cutoff of the 
theory.  This mechanism involves a discrete $\ZZ_N$ symmetry that is non-linearly realized on a scalar as a shift symmetry
and manifests as an exchange symmetry on the $N$ copies of particles with which it interacts~\footnote{Due to the presence of additional discrete symmetries, the $\ZZ_N$ symmetry may be enhanced to a $\DD_N$ symmetry.}.  This approach allows one to obtain
a hierarchy between the mass of the scalar and the UV cutoff that is exponential in $N$, meaning that $N=3$ already solves most problems.

We first consider an explicit example of a Yukawa coupling between a scalar and a fermion.
The starting point is a periodic scalar $\phi$ with period $2 \pi f$.  $\phi$ 
has a spurion $\epsilon$ which breaks the arbitrary shift symmetry $\phi \rightarrow \phi + \alpha$ down to 
$\phi \rightarrow \phi + 2 \pi f$.  Thus $\phi$ only appears in the Lagrangian as
\bea
\epsilon \sin \left ( \frac{\phi}{f} + \theta \right ).
\eea
This structure can result from a perturbative UV completion where $\phi$ is the pseudo-goldstone
of a complex scalar $\Phi \sim e^{i \phi/f}$.

The theory also has a discrete $\ZZ_N$ symmetry under which $\phi$ transforms as 
\bea
\phi \rightarrow \phi + \frac{2 \pi f}{N}.
\eea
It is crucial to the mechanism that the spurion of shift symmetry breaking is not invariant under this symmetry.
Next, we introduce Weyl fermions $\psi_k$ and $\psi_k^c$ for $\phi$ to couple to.  Under the $\ZZ_N$ symmetry, the fermions transform as
\bea
\psi_k , \psi_k^c \rightarrow \psi_{k+1} ,  \psi_{k+1}^c,
\eea  
where we have identified $\psi_0^{(c)}$ and $\psi_N^{(c)}$.
$\phi$ and $\psi$ are coupled in a $\ZZ_N$-symmetric manner via the interaction
\bea
\label{Eq: interact}
\LL = \sum_{k=1}^N \left ( m_\psi + \epsilon \sin \left ( \frac{\phi}{f} + \frac{2 \pi k }{N} \right ) \right )  \psi_k \psi^c_k.
\eea

This Lagrangian describes a scalar $\phi$ coupled to fermions with a Yukawa 
coupling $\sim \epsilon/f$.  Thus the natural 
expectation for the mass of $\phi$ is 
\bea
m_\phi \gtrsim \frac{\epsilon \Lambda}{f}.
\eea
where the UV cutoff is $\Lambda$.
This intuition turns out to be incorrect as the scalar described in Eq.~\ref{Eq: interact} has its potential cancelled to
a very high degree of accuracy.  $N$ insertions of $\epsilon$ will be needed until a non-trivial potential for $\phi$ can be
obtained.

To see the cancelations in action, we take $N > 2$ and consider the potential induced for $\phi$ by closing the fermion loop with two insertions of $\epsilon$:
\bea
V_{1-loop} \supset \sum_{k=1}^N \epsilon^2 \sin^2 \left ( \frac{\phi}{f} + \frac{2 \pi k }{N} \right ) \Lambda^2 = \frac{N}{2} \epsilon^2 \Lambda^2.
\eea
The quadratic divergence to the $\phi$ mass coming from a fermion $\psi_k$ has been cancelled by its $N-1$ partners.
There is a stronger version of the above cancelation, which is that for all integer $m$ with  $N > m \geq 0$,
\bea
\sum_{k=1}^N \sin^m \left ( \frac{\phi}{f} + \frac{2 \pi k }{N} \right ) &=& \begin{cases}
  0 \qquad \qquad \quad \, \, \, m = \text{odd} \\
  \frac{N}{2^m} \frac{m!}{(m/2)!^2} \qquad m = \text{even}
\end{cases}
\eea
Thus the leading-order UV contribution to the potential of $\phi$ comes from the higher-dimensional operator with $N$ insertions of the spurion $\epsilon$.
\bea
V(\phi) &\propto& \frac{\epsilon^N}{\Lambda^{N-4}} \sum_{k=1}^N \sin^N \left ( \frac{\phi}{f} + \frac{2 \pi k }{N} \right )  \\
&=& 
  (-1)^{ \lfloor N/2 \rfloor} \frac{N \epsilon^N}{2^{N-1} \Lambda^{N-4}} \cos \left ( \frac{N \phi}{f} -\frac{\pi}{2} (N \% 2) \right ) .\nonumber
\eea
The UV contribution to the mass of $\phi$ scales as
\bea
m_\phi^2 \sim \frac{N^3 \epsilon^N}{f^2 \Lambda^{N-4}}
\eea
and is exponentially suppressed in $N$.

To get an intuitive understanding of why the potential is so suppressed, we analyze the potential for $\phi$ from a symmetry perspective.  Due to the $\ZZ_N$ symmetry, the potential for $\phi$ must be
$2 \pi f/N$ periodic so that we can express it as
\bea
V(\phi) = \sum_k c_k \sin \left (\frac{N k \phi}{f} + \theta_k \right ) .
\eea
We will see below that under a broad set of assumptions,
\bea
\label{Eq: ci}
c_k \sim \epsilon^{N k}.
\eea
so that by taking $\epsilon$ small and $N$ large, we can parametrically separate the mass of the scalar from the UV cutoff.

The quick and dirty way to derive that $c_k \sim \epsilon^{N k}$ is to note that $e^{i N \phi/f} = ( e^{i \phi/f} )^N$ so that $N$ insertions of the symmetry-breaking spurion $\epsilon e^{i \phi/f}$ are needed to generate a potential for $\phi$.  Alternatively, the relation can be
obtained by considering the potential in frequency space.
The interaction frequency of $\phi$ is $\omega = 1/f$ ($ V \sim \cos(\omega \phi) \psi \psi^c$). 
The frequency associated with the mass term and every other term in the potential for $\phi$ is $N \omega = N/f$ due to the $\ZZ_N$ symmetry.  
In order
to construct the high-frequency mass term, $N$ contributions of the lower frequency $\omega$ are needed.
Each of these comes with its own factor of $\epsilon$ giving the scaling shown in Eq.~\ref{Eq: ci}.

A critical assumption that was made implicitly in the previous discussion is that there are no phase transitions or massless particles as $\phi$ varies in field space.  
Phase transitions introduce discontinuities.  By cutting up a $2 \pi f$ symmetric potential, a $2 \pi f/N$ symmetric potential can
be easily generated.  This point will be exploited later in the paper to obtain a light Higgs boson.

Since $\epsilon$ is a dimension-one number, we need to specify what the dimensionless expansion parameter is.
When dealing with UV contributions to the potential, it is clear that the expansion parameter is
$\epsilon/\Lambda$.  In this case,
\bea
c_k \sim \Lambda^4 \left ( \frac{\epsilon}{\Lambda} \right )^{N k} .
\eea

In addition to UV contributions to the potential, there will also be IR contributions.
As an example of how the IR contributions behave, consider the coupling to the fermions discussed before in Eq.~\ref{Eq: interact}.  The effective potential 
 will depend on the fermion mass $m_\psi$ in some manner, e.g. $m_\psi^4 \log m_\psi$.  From this,
 one sees that the expansion parameter for IR contributions to the potential is $\epsilon/m_\psi$ and that
\bea
c_k \sim \Lambda_{IR}^4 \left ( \frac{\epsilon}{m_\psi} \right )^{N k} .
\eea
Depending on other IR parameters in the Lagrangian, the IR potential for $\phi$ can be relatively unsuppressed.  Thus scalars of this type are sensitive to the IR but insensitive to the UV.

\section{Analytic bounds}

In the example given in the Introduction, we saw that a scalar could couple strongly to 
matter yet remain light.  
In this section, we extend the previous result to more general situations. We describe
under what circumstances the mass of a scalar $\phi$ is exponentially suppressed in $N$
and under what circumstances it is only power-law suppressed in $N$.

Due to the $\ZZ_N$ symmetry, 
the potential for $\phi$ is of the form
\bea
\label{Eq: assumption}
V(\phi) \propto \sum_{k=0}^{N-1} F(\frac{\phi}{f} + \frac{2 \pi k}{N} ).
\eea
Unlike before, we do not assume the existence of a small coupling $\epsilon$.  The results 
presented in this section are valid for any choice of the function $F$ and for any value of $N$ but are most useful in the case where $F$ does not depend explicitly on $N$.

The large-$N$ limit of Eq.~\ref{Eq: assumption} is easy to understand as the sum is simply a Riemann sum that converges to an integral :  
\bea
V(\phi) &\propto& \sum_{k=0}^{N-1} F(\frac{\phi}{f} + \frac{2 \pi k}{N} ) \nonumber \\
\label{Eq: Riemann}
&=& \frac{N}{2 \pi} \int_0^{2\pi} F(\theta) d\theta + \OO(N^0) .
\eea
The leading-order piece is completely $\phi$ independent so that the mass of $\phi$
is subleading in the large-$N$ limit.  Let us denote the subleading piece that generates the
mass of $\phi$ as 
\bea
E_N(F) = \int_0^{2\pi} F(\theta) d\theta - \frac{2 \pi}{N} \sum_{k=0}^{N-1} F(\frac{\phi}{f} + \frac{2 \pi k}{N} ).
\eea
The scaling of the mass of $\phi$ with $N$ is an issue of estimating the convergence rate of 
the Riemann sum of $F''$.  Riemann sums of periodic functions are known to converge extremely 
quickly, and we present two useful theorems (the proofs can be found in Section 9.4 of Ref.~\cite{book}).

The first theorem is a special case of the Euler-Maclaurin theorem.  If the function $F$ is $2 \pi$
periodic and $2m + 1$ times differentiable, then
\bea
| E_N(F) | \leq \frac{2}{N^{2m+1}} \left ( \sum_{j=1}^\infty \frac{1}{j^{2m+1}} \right ) \int_{0}^{2 \pi} | F^{(2 m + 1)} (\theta) | d\theta . \nonumber
\eea
Potentials where there are massless particles as $\phi$ varies will have discontinuities and are not infinitely 
differentiable.  In these cases, the mass of $\phi$ is only power-law suppressed by $N$.

The second theorem can be shown via the residue theorem.  Let $F$ be a $2 \pi$-periodic function 
that is also analytic.  Then there exists an open strip, which includes the real axis and the 
complex axis from $-i a$ to $i a$ with $a > 0$, upon which $F$ can be extended into a holomorphic, $2 \pi$-periodic, bounded function with bound $M$.  
In this case,
\bea
| E_N(F) | \leq \frac{4 \pi M}{e^{N a} - 1} .
\eea
Potentials that are infinitely differentiable give
an exponentially suppressed mass for $\phi$ even if there is no small parameter in the problem.

These two theorems demonstrate how much the mass of $\phi$ can be suppressed for any given
potential.  In the case where there are massless particles as $\phi$ varies, the mass of $\phi$ is suppressed by
how differentiable the potential is.  In the case with no 
discontinuities, the mass is exponentially suppressed even if there is no small number
in the problem.

\section{Examples}

In the example given in the Introduction, we showed how a light scalar (e.g. from a fifth force or dark matter) Yukawa coupled to the SM that naively looks tuned can actually arise naturally.  In this 
section, we provide two more examples of how this new solution to the Hierarchy Problem can be applied to theories of interest.  The first example is that of a light axion.  
This example serves to highlight how different this solution is from other solutions such as supersymmetry, 
which cannot make the axion lighter than its QCD contributions.
The second example is a solution to the Little Hierarchy Problem.

\subsection{A light axion}

Unlike the example in the Introduction, the axion~\cite{Peccei:1977hh,Peccei:1977ur,Weinberg:1977ma,Wilczek:1977pj} does not have a small parameter $\epsilon$ characterizing its couplings.  
However, the axion potential is analytic so that by the
theorems presented before, its mass must be exponentially suppressed in the large-$N$ limit.
As before, we have a $\ZZ_N$ symmetry with $N$ copies of the SM that are interchanged under the 
symmetry, and an axion that non-linearly realizes the discrete symmetry.
The axion couples to the $N$ different sectors via the coupling
\bea
\LL = \sum_k \left ( \frac{a}{f} + \frac{2 \pi k}{N} + \theta \right ) G_k \tilde G_k .
\eea

Due to confinement of the $N$ different QCD sectors, there will be a potential for the axion whose leading order contribution is
\bea
V(a) = &-& m_\pi^2 f_\pi^2 \sum_k \sqrt{1 - 4 \frac{m_u m_d}{(m_u + m_d)^2} \sin^2 \left ( \frac{a}{2f} + \frac{\pi k}{N} \right ) } \nonumber \\
\label{Eq: axion}
&+& \OO(m_\pi^4).
\eea
As the theta angles are all identical, we have shifted them away.
With a bit of algebra, one can show that $\theta = 2 \pi k/N$ for integer $k$ is a minimum (maximum) for $N$ odd (even).

To use the results of the convergence theorems presented before, we define
\bea
F(z) = \sqrt{1 - 4 \frac{m_u m_d}{(m_u + m_d)^2} \sin^2 \left ( \frac{z}{2} \right ) } ,
\eea
which is holomorphic until the square root of a negative number is taken.  We are thus considering
the open strip $(-i a , i a)$ where 
\bea
a = \log ( c + \sqrt{c^2 - 1} ), \qquad c = \frac{(m_u + m_d)^2}{2 m_u m_d} - 1 .
\eea
Inserting the measured values of the quark masses, we find that the mass of the axion is bounded to be $\approx 1/2^{N/2}$ smaller than its natural value where we have neglected the subleading terms proportional to powers of $N$.

Eq.~\ref{Eq: axion} contains subleading terms suppressed by more powers of the quark masses.  
These terms are also analytic and thus also give exponentially suppressed contributions to the axion mass.  
These contributions are also suppressed by at least $2^{N/2}$.  To understand this scaling behavior,
we note that if $m_u = m_d$, there is a first-order phase transition at $\theta = \pi$~\cite{Witten:1980sp}.  If the quark masses are equal, the axion
potential is non-analytic and thus not exponentially suppressed.  Any non-zero mass difference
results in an analytic potential, and the parameter governing the exponential suppression is necessarily $m_u/m_d \sim 1/2$.
As the exponential suppression of the mass is due to how far away the potential is from discontinuities, a.k.a. phase transitions, the axion mass
is necessarily suppressed by at least $1/2^{N/2}$.

Aside from analytically bounding the mass of the axion in the large-$N$ limit, we also numerically fit the exponential dependence of the axion mass on $N$ and find that
\bea
\frac{m_a(N)}{m_a(N=1)} \sim \frac{4}{2^{N/2}}
\eea
There is good agreement between the analytically-derived limit on the mass and actual mass.

The price of obtaining an exponentially lighter axion is the linear problem associated with the fact that only one of the $N$ copies of the SM has $\theta = 0$.  The rest have $\theta = 2 \pi k/N$.  Thus one has traded an exponential fine tuning in the mass for a linear tuning of why we are in the sector with $\theta = 0$.  This secondary problem may be solvable via other mechanisms.

\subsection{A naturally light Higgs boson} \label{Sec: Higgs}

Solving the Hierarchy Problem is fundamentally about finding a reason that a Higgs mass of zero is special.  
The scalars discussed in this article are sensitive to phase transitions.  As a phase transition occurs when the Higgs mass crosses zero,
these scalars should be able to favor a small Higgs mass.

Following this train of thought, in this subsection we develop a theory for the modulus of the Higgs mass.  
A modulus coupling to $N=3$ or $4$ copies of the Higgs bosons can result in one of them
being lighter than what naturalness would otherwise imply by a factor of 10, thereby solving the Little Hierarchy Problem.
If the other $N-1$ copies of the Higgs boson had positive masses, then there would only need to be N copies of the Higgs boson with the rest of the SM transforming trivially under the $\ZZ_N$.
In the model presented below, the other Higgs bosons obtain a negative mass squared so that the entire SM needs to be copied.

\subsubsection{$N=3$/$N=4$ case}

Consider a $\mathbb{Z}_N$ symmetry under which there are $N$ copies of the SM
and a scalar $\phi$, which is the modulus of the Higgs mass~\footnote{In principle, $SU(3)_c \times U(1)_Y$ could transform trivially under this exchange symmetry,
but the resulting light colored and charged particles have been excluded by experiment.}.  For the rest of this section,
$N$ will be 3 or 4.  
We couple $\phi$ to the Higgs with a shift symmetry-breaking parameter $\epsilon^2$:
\bea
\label{Eq: Higgs}
V &=& \sum_k m_{H,k}^2(\phi) H_k H^\dagger_k + \lambda (H_k H^\dagger_k)^2 ,\\
\nonumber m_{H,k}^2(\phi) &=& -m_H^2 + \epsilon^2 \cos \left ( \frac{\phi}{f} + \frac{2 \pi k}{N} \right ) .
\eea
For simplicity, we will take all cross-quartic couplings between the Higgses to be zero, but our results will not depend
on this assumption.  We will take $m_H^2 > 0$ and 
$\Lambda^2 > \epsilon^2 \gtrsim m_H^2 = 3 y_t^2 \Lambda^2/8 \pi^2$.  As discussed before, the UV contribution to the $\phi$ potential
will be suppressed by $\epsilon^{2N}/\Lambda^{2N-4}$.

\begin{figure}
  \centering
  \includegraphics[width=0.45\textwidth]{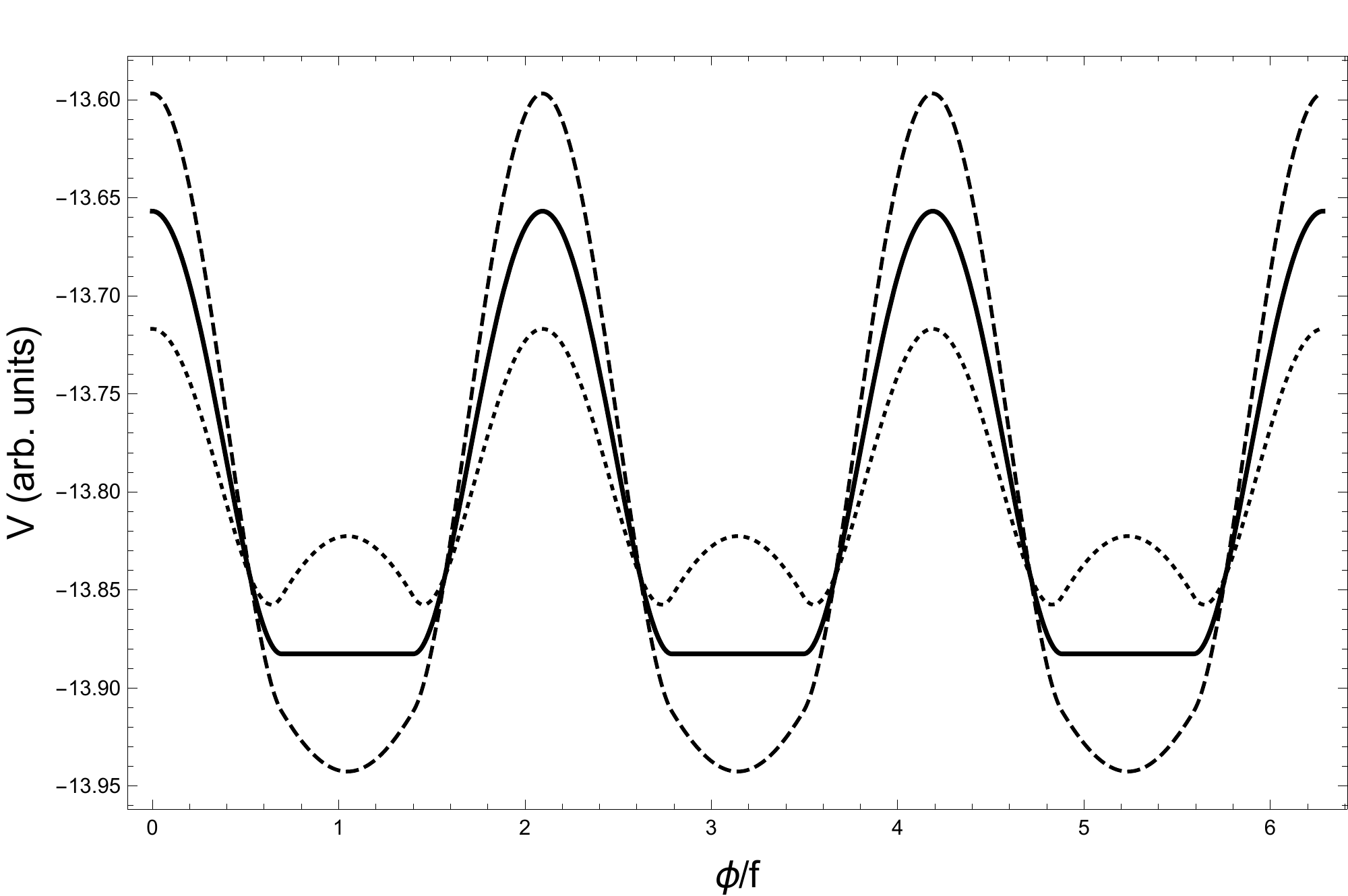}
  \caption{ The $N=3$ tree-level potential when $\epsilon^2 = 1.3 \, m_H^2$.  The solid line is the potential with no UV contribution, while the dotted (dashed) line includes the UV generated potential with $\theta = 0$ ($\theta = \pi$) in Eq.~\ref{Eq: potential}.  If the UV-generated potential has a minimum where the Higgs masses are all negative, then all Higgs masses are at their large natural value.  If the UV-generated potential has a minimum where one of the Higgs masses is positive, then there is a light Higgs.
    \label{fig}}
\end{figure}

The Higgs naturally has a phase transition when its mass changes sign.  Thus the potential for $\phi$ will be sensitive to changes in
the sign of the Higgs mass.  To see this effect explicitly, we
integrate out the Higgs classically.  Only if the total Higgs mass is negative will the Higgs induce a non-zero tree-level potential for 
$\phi$.  The tree-level potential for $\phi$ is
\bea
V = &-& \frac{\epsilon^{2N}}{\Lambda^{2N-4}} \cos \left( N \frac{\phi}{f} + \theta \right )  \nonumber \\
\label{Eq: potential}
&-& \sum_k \frac{m_{H,k}^4(\phi)}{4 \lambda} \quad \Theta \left ( -m_{H,k}^2(\phi) \right ) .
\eea
By previous arguments, if all three Higgs masses are negative~\footnote{Requiring that all three Higgs masses are negative for certain values of $\phi$ and that one of them becomes positive for other values of $\phi$ corresponds to the choice that
$2 m_H^2 > \epsilon^2 > m_H^2$.},
 then the contribution to the potential from the Higgs is $\phi$ independent.  However, as soon as some of the Higgs masses become positive, a phase transition occurs and there is a potential for $\phi$.

An example $N=3$ potential is shown in Fig.~\ref{fig} for some specific choices of parameters.
The preference for small Higgs masses can be seen by considering the Higgs's contribution to the
potential of $\phi$.  
Over some of parameter space, all three Higgs vevs are negative and $\phi$ does not acquire a potential 
from the Higgses.  However, whenever one of the Higgs masses becomes positive, there is no longer 
a cancelation and the potential quickly increases.  Thus this contribution to the potential has a 
minimum whenever all of the Higgs masses are negative.
This preference for negative Higgs masses is balanced against the $\epsilon$-suppressed UV contribution to the potential.  Choosing the phase of the UV contribution to favor positive Higgs masses gives a theory where at the minimum of the potential, one of the sectors has a Higgs with a small positive mass.

\begin{figure}
  \centering
  \includegraphics[width=0.45\textwidth]{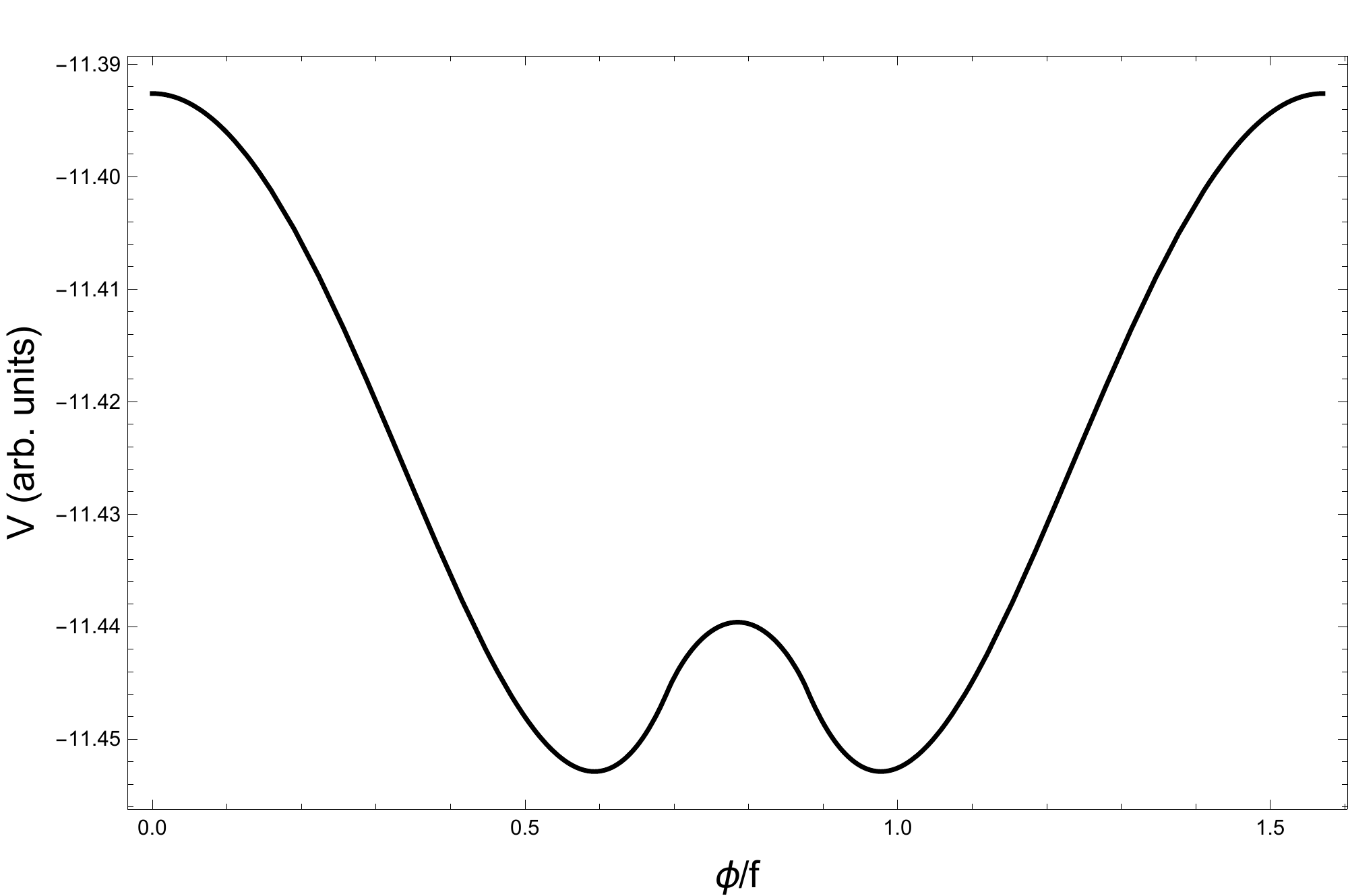}
  \caption{ The $N=4$ potential including 1-loop effects for $\epsilon^2 = 1.3 \, m_H^2$ and $\beta = 0.2$.  At the minimum, one of the four Higgs masses is $\sim 0.1 \, m_H$ and is positive.  As the Higgs is a factor of $\sim 10$ lighter than its natural value, it solves the Little Hierarchy Problem modulo getting the sign of the Higgs mass wrong.
    \label{fig2}}
\end{figure}

When $N \gtrsim 3$, the UV contribution becomes subdominant to the 1-loop potential for $\phi$.
The 1-loop Coleman-Weinberg potential gives a potential for $\phi$ that is of the form
\bea
V_\text{1-loop} = \frac{\beta}{16 \pi^2} \sum_k \left ( H_k H_k^{\dagger} \right )^2 \log H_k H_k^{\dagger} /m_H^2.
\eea
The sign and value of $\beta$ is determined by the beta functions at the natural scale of the Higgs masses.
The $N=4$ potential including the 1-loop potential ($\beta = 0.2$) is shown in Fig.~\ref{fig2}.

The previous two examples of $N=3$ and $N=4$ gave a small positive Higgs mass, as opposed to the observed small negative Higgs mass.  There are two simple ways of obtaining a small negative Higgs mass.  The first is to introduce a small amplitude but high-frequency sine wave
potential for $\phi$,
\bea
V = \alpha \cos \left ( \frac{M \phi}{f} \right ) .
\eea
This will introduce additional minima, but can result in a small negative Higgs
mass  in the {\it absolute} minimum.  For example, including the 1-loop potential for $N=4$, the
values $\alpha = -0.01 \, m_H^4$, $M=36$ and $\beta = 0.1$ give a negative Higgs mass at the absolute minimum.
For this parameter set, the UV cutoff is 10 TeV.

Another way of obtaining a small negative Higgs mass is to introduce an additional $\ZZ_2$ symmetry under which the $N$ copies of the SM are taken to another $N$ copies.  In the limit
of an exact $\ZZ_2$ symmetry, there are two light Higgses with identical small positive masses.  
The $\ZZ_2$ symmetry is softly broken by giving the two sectors slightly different Higgs masses, $m_H^2$ and $m_H^2 + \delta^2$.  A small negative Higgs mass can result if $\delta \gtrsim 125$ GeV.
For example, in the $N=4$ case above, a small $\delta^2 \sim 0.02 m_H^2$ and $\beta = 0.05$ results in two light Higgses each a factor of 10 lighter than the bare mass.  One of the two has a positive mass squared while the other has a negative mass squared.

Taking the large-$N$ limit of the above solution to the Little Hierarchy Problem does not produce any parametric
enhancements, though there are small numerical benefits going beyond $N=3$ or $4$.  The main issue is that requiring that there is only one Higgs mass that is positive
at a time means that $m_H^2/\cos \left ( \pi/N \right ) > \epsilon^2 > m_H^2$.  In the large-$N$ limit, this amounts to a $1/N^2$ tuning.
Even accepting this tuning, the other major issue is that the 1-loop Coleman-Weinberg potential is not parametrically suppressed.  By explicit calculation, the Coleman-Weinberg potential is only $1/N^2$ suppressed, which does not result in a small enough Higgs mass.

Many intriguing possibilities arise if the Coleman-Weinberg potential were highly suppressed due to very small beta functions.  Taking $N \sim 10$ would suppress the UV contributions to a point where this would be a solution to the full Hierarchy problem.  Additionally, the minimum with a negative Higgs mass near 100 GeV could be generated by features in the beta function, e.g. a small increase and decrease in the quartic around 100 GeV.  This offers a unique twist on how conformal dynamics might result in a light Higgs mass.

\subsubsection{Phenomenology}

The phenomenology of this solution to the Little Hierarchy Problem will be discussed in detail in future 
work.  Here we summarize the salient features.

Preferential reheating of the sector with a light negative Higgs mass is built into the model as $\phi$ couples to the SM 
Higgses through scalar mixing and is exactly the scalar reheating model described in 
NNaturalness~\cite{Arkani-Hamed:2016rle}.  Since $\phi$ is naturally lighter than the lightest Higgs, its decays
preferentially reheat the sector with the lightest Higgs, evading all current cosmological 
constraints.  Thus if $\phi$ mediates reheating to the SM, then all cosmological problems are 
naturally avoided.

Another feature of this solution is that the only particle required to interact with the Higgs is $\phi$.  Much like the axion solution to the 
strong CP problem, this mechanism works for any value of $f$ and a large value of $f$ results in $\phi$ being very difficult to detect.  
In this limit, $\phi$ shares all of the same
benefits and problems as the axion, e.g. care is needed so as not to overclose the universe, but on the flip side, $\phi$ can provide a dark matter candidate.

The details will depend on the particular models, but parametrically $m_\phi \sim \text{TeV}^2/f$ and the mixing with the Higgs scales as $\sim 10$ TeV/$f$.
$\phi$ with masses down to $\sim$ 0.1 GeV ($f \lesssim 10^7$ GeV) are excluded by meson decays and beam dumps (see Ref.~\cite{Alekhin:2015byh,Flacke:2016szy} for a compendium of constraints).
Horizontal branch star cooling constrains scalar couplings to electrons and excludes $\phi$ in the range $10^{13}$ GeV $\gtrsim f \gtrsim 10^{10}$ GeV, while fifth-force experiments exclude $f \gtrsim 10^{17}$ GeV (see Ref.~\cite{Raffelt:2012sp,Hardy:2016kme} for a compendium of constraints).  These estimates are very rough and detailed constraints will be model dependent.

Finally, if there is a non-zero cross-quartic coupling between the Higgses, then the Higgs can mix with
the other Higgses.  Due to the requirement of vacuum stability and the relatively small value of the SM quartic coupling,
negative cross-quartics larger than a few percent are excluded.  Positive cross-quartics larger than a few percent are
also excluded, as large cross-quartics generally push the theory out of the parameter space where all three Higgses
can obtain vevs.  As a result, the mixing between the multiple Higgses is suppressed by $\approx$ few $\times \, 10^{-3}$.
The resulting exotic collider signatures are very similar to Twin Higgs models (see e.g. Refs.~\cite{Burdman:2014zta,Craig:2015pha}) only with much smaller production rates.
Another difference is the absence of the $v/f$ tuning needed to make electroweak symmetry breaking
work in Twin Higgs models.

\section{Conclusion}

To conclude, we briefly compare our new solution to similar solutions to the Hierarchy Problem,
Twin Higgs~\cite{Chacko:2005pe} and Little Higgs~\cite{ArkaniHamed:2001nc}.  In unitary gauge, both Little Higgs and Twin
Higgs solve the gauge and Yukawa divergences by coupling the Higgs as $\sin v/f$ to our sector and as $\cos v/f$
to our partners.  The cancelations are then just a $\ZZ_4$ version of the previous arguments, where two copies
have been removed because gauge invariance cancels the odd powers of $v$ so that the extra two copies are not needed
for the cancelation.  This discrete symmetry solution utilizes a generalization of these sin/cos identities though in a completely different manner.

The approach most similar to the one presented in this paper are dimensional deconstructions of an extra dimension where the scalars 
only pick up a mass non-locally~\cite{ArkaniHamed:2001ca,ArkaniHamed:2001nc}.  The dimensional deconstruction based theories result in effective field theories that are a 
subset of our more general approach.  In that language, the scalar is light due to 5D gauge invariance and locality.
In contrast, as can be seen from our discrete symmetry based approach, the $N$ sectors do not need to be Higgsed down to the 
diagonal subgroup (no fifth dimension is required) and no sense of locality is needed in
theory space.  Any interaction is allowed as long as the interactions satisfy the $\ZZ_N$ symmetry.

The most obvious extension for this approach is to use it on the Higgs boson directly.  There are two challenges for this approach.  The first is that our solution to the Hierarchy Problem suppresses the Higgs quartic.
The second is that while gauge charged scalars can be made compact, they cannot be made 
periodic.  If there were a non-abelian equivalent  of frequency, then there might be a way to solve the quadratic divergences coming from the Yukawa 
couplings in this manner.

There is still much to be explored with this new solution to the Hierarchy Problem.  We briefly list a few below:
\begin{itemize}
\item The $\ZZ_N$ solutions to the Little Hierarchy Problem need to be 
explored both theoretically and phenomenologically.  Ideally there exists a model where the only particles transforming under the $\ZZ_N$ symmetry are the Higgs boson and the modulus.  In this case, the top quark would be its own partner.
\item UV completions of these theories would also be interesting, since moduli frequently appear in String Theory and very often have discrete symmetries associated with them.
\item Many theories of flavor involve discrete symmetries, hence a scalar that realizes the discrete symmetry non-linearly 
may allow for interesting flavor physics and possibly even explain why the universe has three generations.
\item This new solutions allows for fifth forces and scalar dark matter that would otherwise appear tuned.
\item The moduli can be dark matter, leading to new and interesting signatures in the early universe and at experiments.
\item This mechanism allows for the inflaton to have large couplings and not ruin its flat potential.
\end{itemize}

Very optimistically, the sensitivity of this scalar to phase transitions
leads one to hope that this approach could help solve the cosmological constant problem as well.
At the very least, the recent spate of solutions to the Hierarchy Problem~\cite{Graham:2015cka,Arkani-Hamed:2016rle} demonstrates that there is still much to be learned about naturalness.

\section*{Acknowledgements}

A.H. is supported by NSF Grant PHY-1620074, DOE Grant DE- SC0012012 and by the Maryland Center for Fundamental Physics (MCFP).  A.H. thanks Savas Dimopolous, Junwu Huang, and Raman Sundrum for useful discussions.

\bibliography{reference}

\end{document}